\documentclass[useAMS,usenatbib,usegraphicx]{mn2e}
\usepackage{epsf}
\usepackage{longtable}

\newcommand{\simgt}{\lower.5ex\hbox{$\; \buildrel > \over \sim \;$}}
\newcommand{\simlt}{\lower.5ex\hbox{$\; \buildrel < \over \sim \;$}}

\title[Properties of SN Refsdal]
{Predicted properties of multiple images of the strongly lensed
  supernova SN Refsdal}

\author[M.~Oguri]
{Masamune Oguri,$^{1,2,3}$\thanks{E-mail: masamune.oguri@ipmu.jp} 
\\
$^1$Research Center for the Early Universe, University of Tokyo, 
7-3-1 Hongo, Bunkyo-ku, Tokyo 113-0033, Japan\\
$^2$Department of Physics, University of Tokyo, 7-3-1 Hongo,
Bunkyo-ku, Tokyo 113-0033, Japan\\
$^3$Kavli Institute for the Physics and Mathematics of the Universe
(Kavli IPMU, WPI), University of Tokyo, Chiba 277-8583, Japan\\
}

\begin{document}

\date{\today}

\voffset- .5in

\pagerange{\pageref{firstpage}--\pageref{lastpage}} \pubyear{}

\maketitle

\label{firstpage}

\begin{abstract}
We construct a mass model of the cluster MACS J1149.6+2223 to study
the expected properties of multiple images of SN Refsdal, the first
example of a gravitationally lensed supernova with resolved multiple
images recently reported by Kelly et al. We find that the best-fit
model predicts six supernova images in total, i.e., two extra images
in addition to the observed four Einstein cross supernova images
S1--S4. One extra image is predicted to have appeared about 17 years
ago, whereas the other extra image is predicted to appear in about one
year  from the appearance of S1--S4, which is a testable prediction
with near future observations.  The predicted magnification factors of
individual supernova images range from $\sim 18$ for the brightest
image to $\sim 4$ for the faint extra images. Confronting these
predictions with future observations should provide an unprecedented
opportunity to improve our understanding of cluster mass
distributions.  
\end{abstract}


%%%%%%%%%%%%%%%%%%%%%%%%%%%%%%%%%%%%%%%%%%%%%%%%%%%%%%%%%%%%%
%%%%%%%%%%%%%%%%%%%%%%%%%%%%%%%%%%%%%%%%%%%%%%%%%%%%%%%%%%%%%
\section{Introduction}
%%%%%%%%%%%%%%%%%%%%%%%%%%%%%%%%%%%%%%%%%%%%%%%%%%%%%%%%%%%%%
%%%%%%%%%%%%%%%%%%%%%%%%%%%%%%%%%%%%%%%%%%%%%%%%%%%%%%%%%%%%%

The existence of strong gravitationally lensed supernovae has
long been predicted \citep[e.g.,][]{holz01,goobar03,oguri03a,oguri10b}.
They provide a unique opportunity to study the mass distribution of
lensing objects as well as cosmological parameters. For instance, the
transient nature of supernovae makes it straightforward to measure
arrival time differences of multiply imaged supernovae, which enables
us to directly measure the absolute distance scale of the Universe 
\citep{refsdal64}. Direct measurements of magnification factors are
also possible if Type Ia supernovae are gravitationally lensed, by
taking advantage of their standard candle nature. The magnification
factor contains valuable information that leads to breaking
degeneracies in lens potentials and cosmological parameters
\citep{kolatt98,oguri03b,bolton03}. 

Gravitationally lensed supernovae are predicted to be rare, which is
why they have not been explored until recently. While several strongly 
magnified supernovae have been discovered behind massive clusters
\citep{amanullah11,patel14,nordin14}, multiple images were not
produced for these events. \citet{quimby13,quimby14} showed that the
unusual transient PS1-10afx \citep{chornock13} at $z=1.389$ was in fact
a gravitationally lensed Type Ia supernova, in which PS1-10afx was
shown to be magnified by a factor of 30 by a foreground lensing galaxy
at $z=1.117$. Its large magnification factor suggests that
PS1-10afx must be multiply imaged (with most likely four images),
which indicates that it is presumably the first discovery of a
strongly lensed supernova with multiple images. However,
the angular resolution of the Pan-STARRS survey was not high enough to
resolve the multiple images of PS1-10afx, and hence no time delay
measurement was obtained. 

Recently, \citet{kelly15} reported the first example of {\it resolved}
multiple images of a strongly lensed supernova from the Grism
Lens-Amplified Survey from Space (GLASS; PI: T. Treu). The supernova
`SN Refsdal' is located in a spiral galaxy at $z=1.491$ which itself
is multiply imaged by the cluster MACS J1149.6+2223 at
$z=0.544$. \citet{kelly15} identified four supernova images around an
elliptical cluster member galaxy, which constitute a cross-like image
configuration.  

In this {\it Letter}, we present an initial attempt to predict the
property of the multiple images of SN Refsdal from a currently
available lens mass model, which is crucial given the timeliness of
this event. We assume a flat universe with matter
density $\Omega_M=0.26$, cosmological constant  $\Omega_\Lambda=0.74$, 
and Hubble constant $H_0=72\,{\rm km\,s^{-1}Mpc^{-1}}$. 
The J2000 coordinates are used throughout the paper.

%%%%%%%%%%%%%%%%%%%%%%%%%%%%%%%%%%%%%%%%%%%%%%%%%%%%%%%%%%%%%
%%%%%%%%%%%%%%%%%%%%%%%%%%%%%%%%%%%%%%%%%%%%%%%%%%%%%%%%%%%%%
\section{Mass Modelling of MACS J1149.6+2223}\label{sec:model}
%%%%%%%%%%%%%%%%%%%%%%%%%%%%%%%%%%%%%%%%%%%%%%%%%%%%%%%%%%%%%
%%%%%%%%%%%%%%%%%%%%%%%%%%%%%%%%%%%%%%%%%%%%%%%%%%%%%%%%%%%%%

%%%%%%%%%%%%%%%%%%%%%%%%%%%%%%%%%%%%%%%%%%%%%%%%%%%%%%%%%%%%%
\subsection{Data}
%%%%%%%%%%%%%%%%%%%%%%%%%%%%%%%%%%%%%%%%%%%%%%%%%%%%%%%%%%%%%

The cluster MACS J1149.6+2223 was observed with the {\it Hubble Space
  Telescope (HST)} in several different programmes. Early {\it HST}
observations revealed the presence of a multiply imaged face-on spiral
galaxy at $z=1.491$ \citep{zitrin09,smith09}. The cluster is also
included in the ongoing Frontier Fields programme (FF; PI: Lotz). In
this paper we use `pre-FF {\it HST} imaging' data provided in the FF
website\footnote{http://archive.stsci.edu/prepds/frontier/lensmodels/},
which were obtained in the Cluster Lensing and Supernova Survey with
Hubble \citep[CLASH;][]{postman12}. 
We select cluster member galaxies from the red-sequence in the
colour-magnitude diagram with F475W and F814W band images, where
galaxy magnitudes are extracted with SExtractor \citep{bertin96}. For
strong lensing constraints, we again use the same constrained used for
constructing `pre-FF' mass models  \citep[see,
  e.g.,][]{johnson14,richard14}. More specifically we adopt 
the multiple image sets ID 1--8 and ID 13 presented in
\citet{johnson14} for our lens mass modelling. We note that ID 9 and
10 are excluded because they are located far from the cluster centre, 
and ID 14 because its constraining power is weak given only two
images. Spectroscopic redshifts are available only for ID 1--3; for
multiple image sets with unknown source redshifts, we treat their
redshifts as free parameters.

We can also take advantage of detailed morphological features of the
$z=1.491$ lensed spiral galaxy \citep[see also][]{smith09}.
In this paper, we identify 4 additional sub-peaks, in addition to the
main core of the spiral galaxy, in the spiral arm in the {\it HST}
image. We use the preliminary mass model constructed without including
the sub-peaks to match these morphological features between different
spiral galaxy images. The identified and cross-matched features are
then included as observational constraints to further refine the mass
model. We note that our identifications and grouping of the features
are in accord with those presented in \citet{smith09} and
\citet{rau14}. 

%%%%%%%%%%%%%%%%%%%%%%%%%%%%%%%%%%%%%%%%%%%%%%%%%%%%%%%%%%%%%
\subsection{Mass Modelling}
%%%%%%%%%%%%%%%%%%%%%%%%%%%%%%%%%%%%%%%%%%%%%%%%%%%%%%%%%%%%%

We use the public software {\it glafic} \citep{oguri10a} to construct
the mass model of MACS J1149.6+2223. In short, {\it glafic} adopts a
parametric modelling approach in which the lens potential is
described by a sum of multiple components describing dark matter
haloes, cluster member galaxies, and various perturbations, and the
model parameters are optimized to match observed positions of multiply
imaged objects. 

We describe a dark halo component by an elliptical extension of the
so-called NFW \citep{navarro97} density profile. The ellipticity is
introduced in the isodensity contour of the convergence. The main NFW
halo is placed at the centre of the brightest cluster galaxy (BCG) at 
(R.A., Decl.)=(177.398749, 22.398531). In addition to the main halo,
we place two additional haloes at the centres of bright cluster member
galaxies at (177.392933, 22.411875) and (177.406453, 22.389575). We
leave the ellipticity and position angle of the main halo at the BCG
as free parameters, whereas ellipticities and position angles of the
two sub-halo components are fixed to those of the observed cluster
galaxies (see below). In all the halo components, the halo mass and
concentration parameter are treated as free parameters.

Member galaxy components are modelled by pseudo-Jaffe ellipsoids,
which contain the velocity dispersion $\sigma$ and the truncation
radius $r_{\rm trunc}$ as model parameters. In order to reduce the
number of parameters, we adopt the scaling of these parameters with
the galaxy luminosity in F814W band, $\sigma \propto L^{1/4}$ and
$r_{\rm trunc} \propto L^{1/2}$ 
\citep[the constant mass-to-light ratio;][]{oguri10a},
and regard the overall normalizations
of the scaling relations as free parameters. The ellipticity and
position angles of individual member galaxies are fixed to values
measured in the F814W image by SExtractor.
To further improve the modelling accuracy we also include small
perturbations of the form $\phi \propto r^2\cos m\theta$ in the lens
potential. Specifically we include $m=2$ and $3$ terms. For each term,
perturbation amplitude and position angle are model parameters.

The cluster member galaxy at (177.397782, 22.395448) that produces the
four Einstein cross supernova images (S1--S4) has a critical
importance for fitting the supernova images. Thus we do not use the
scaling relation mentioned above for this particular galaxy. Instead 
we place a separate pseudo-Jaffe ellipsoid with all the parameters
($\sigma$, $r_{\rm trunc}$, ellipticity, and position angle) as free
parameters. We use the positions of identified multiple images as
observational constraints. 
 We follow the standard mass modelling procedure to adopt
  positional errors larger than measurement errors considering effects
  of complexities of mass distributions such as asymmetry and
  substructures which are not fully taken into account in our
  parametric mass modelling. For most multiple images we adopt the
positional uncertainty (in the image plane) of $\sigma=0\farcs4$ which
is a typical accuracy of recent cluster mass modelling
\citep[e.g.,][]{ishigaki15} and is also supported by the $\chi^2$
value of our best-fitting model as we will see below.
Since the lensed spiral galaxy is the
host galaxy of the lensed supernova, we require slightly better
accuracy of $0\farcs2$ for the main core and the 4 sub-peak images of
the spiral galaxy. Finally we also include the observed positions of
the lensed supernova \citep{kelly15}, which were also derived using
the world coordinate system of the pre-FF {\it HST} imaging data,  
assuming even better positional uncertainty of $0\farcs05$ as the
accurate reproduction of supernova image positions is important for
the accurate prediction of time delays.  In total 46 images from 14
sources are used for mass modelling. On the other hand we have 52
model parameters, indicating that the number of the degree of freedom
of $\chi^2$ fitting is 40.  

%%%%%%%%%%%%%%%%%%%%%%%%%%%%%%%%%%%%%%%%%%%%%%%%%%%%%%%%%%%%%%%%%%%%
\begin{figure*}
\begin{center}
 \includegraphics[width=0.8\hsize]{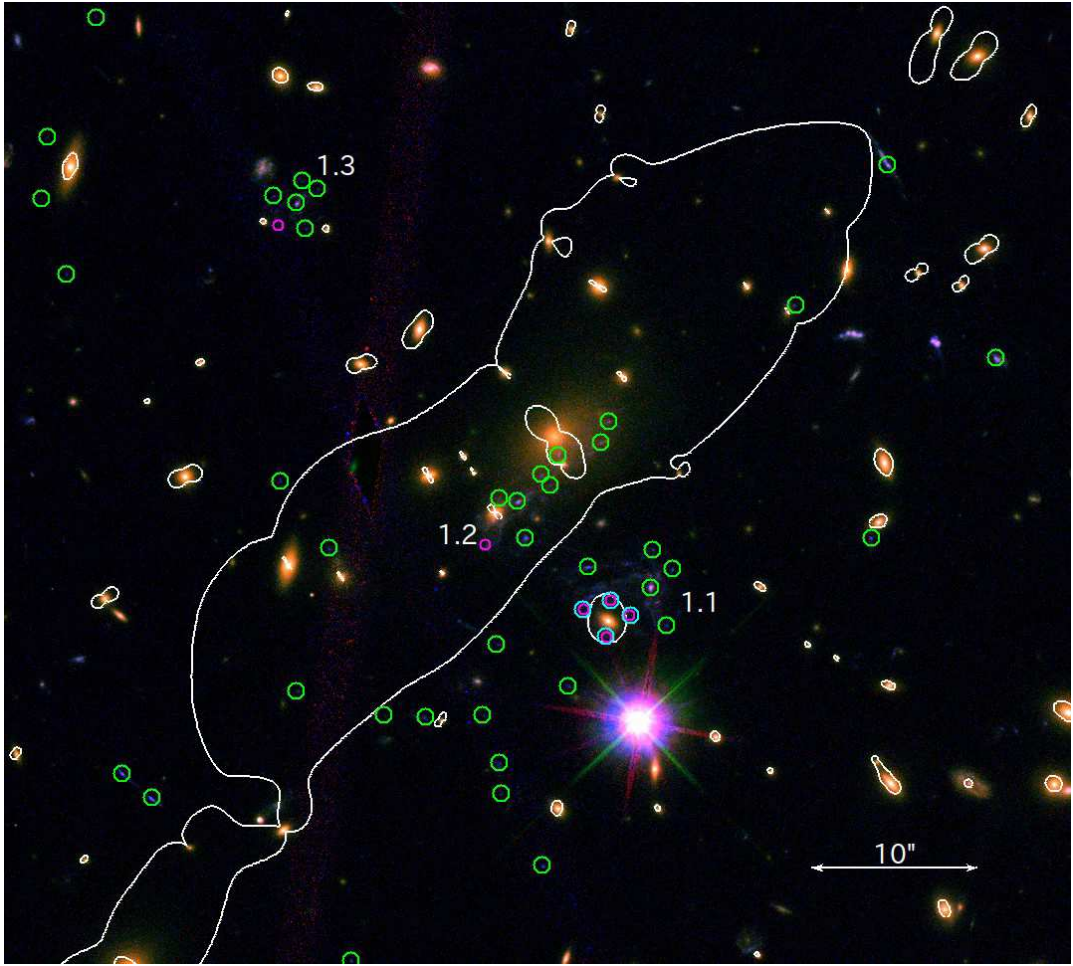}
\end{center}
\caption{The colour-composite image of MACS J1149.6+2223 from the {\it
    HST} F475W-, F814W-, and F850LP-band images. The solid curve shows
  critical curves at $z=1.491$ predicted by our best-fit model. 
  North is up and West is left. Circles indicate observed multiple
images (galaxy images with green circles and supernova images with 
cyan circles) that are used for mass modelling. The regions marked by
`1.1', '1.2', and `1.3' are the locations of multiple images of a
face-on spiral galaxy at $z=1.491$, in which SN Refsdal exploded.
Predicted positions of six multiple images of SN Refsdal are marked by
small magenta circles.
\label{fig:main}}
\end{figure*}
%%%%%%%%%%%%%%%%%%%%%%%%%%%%%%%%%%%%%%%%%%%%%%%%%%%%%%%%%%%%%%%%%%%%%

%%%%%%%%%%%%%%%%%%%%%%%%%%%%%%%%%%%%%%%%%%%%%%%%%%%%%%%%%%%%%%%%%%%%
\begin{figure*}
\begin{center}
 \includegraphics[width=0.32\hsize]{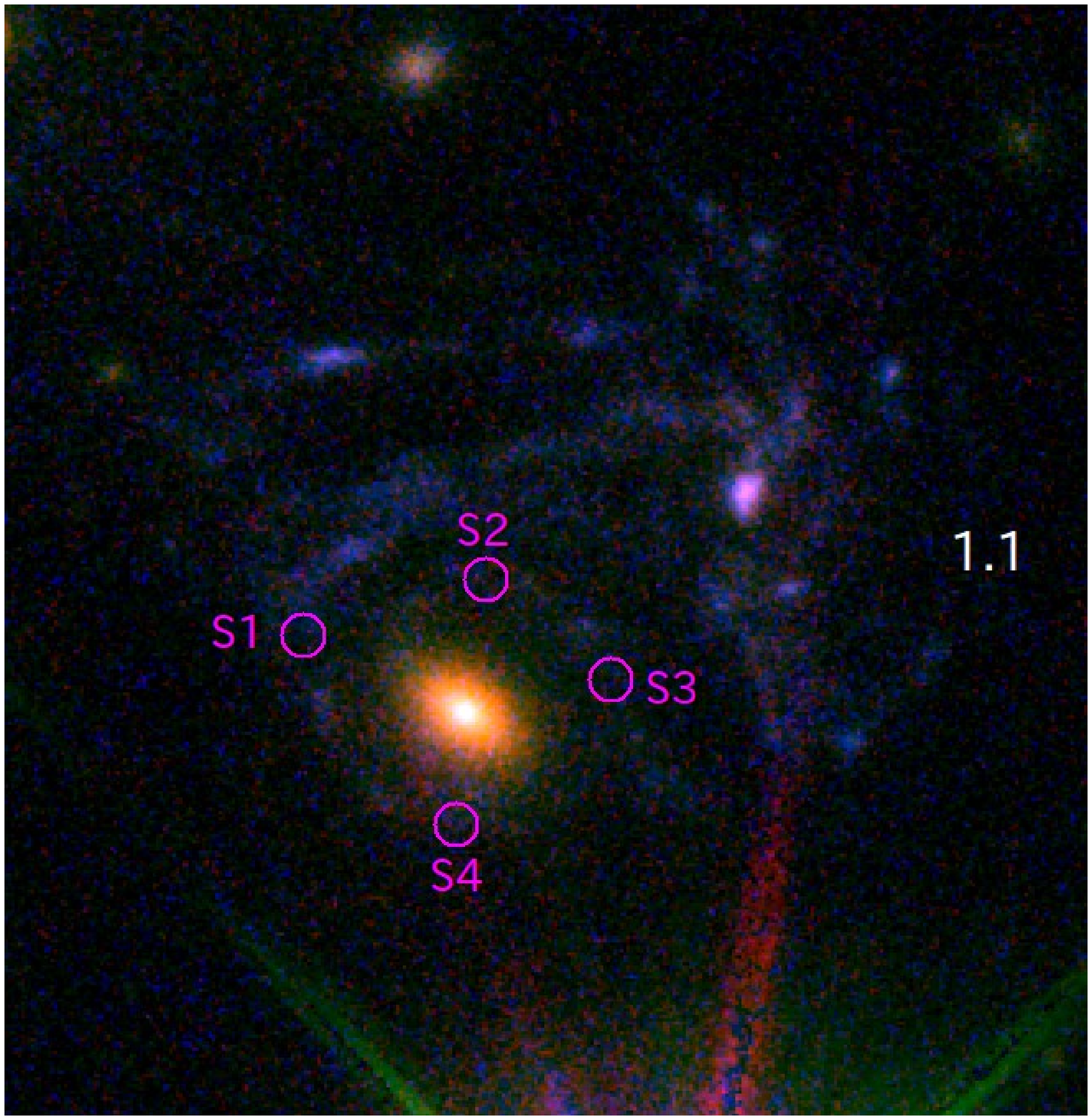}
 \includegraphics[width=0.32\hsize]{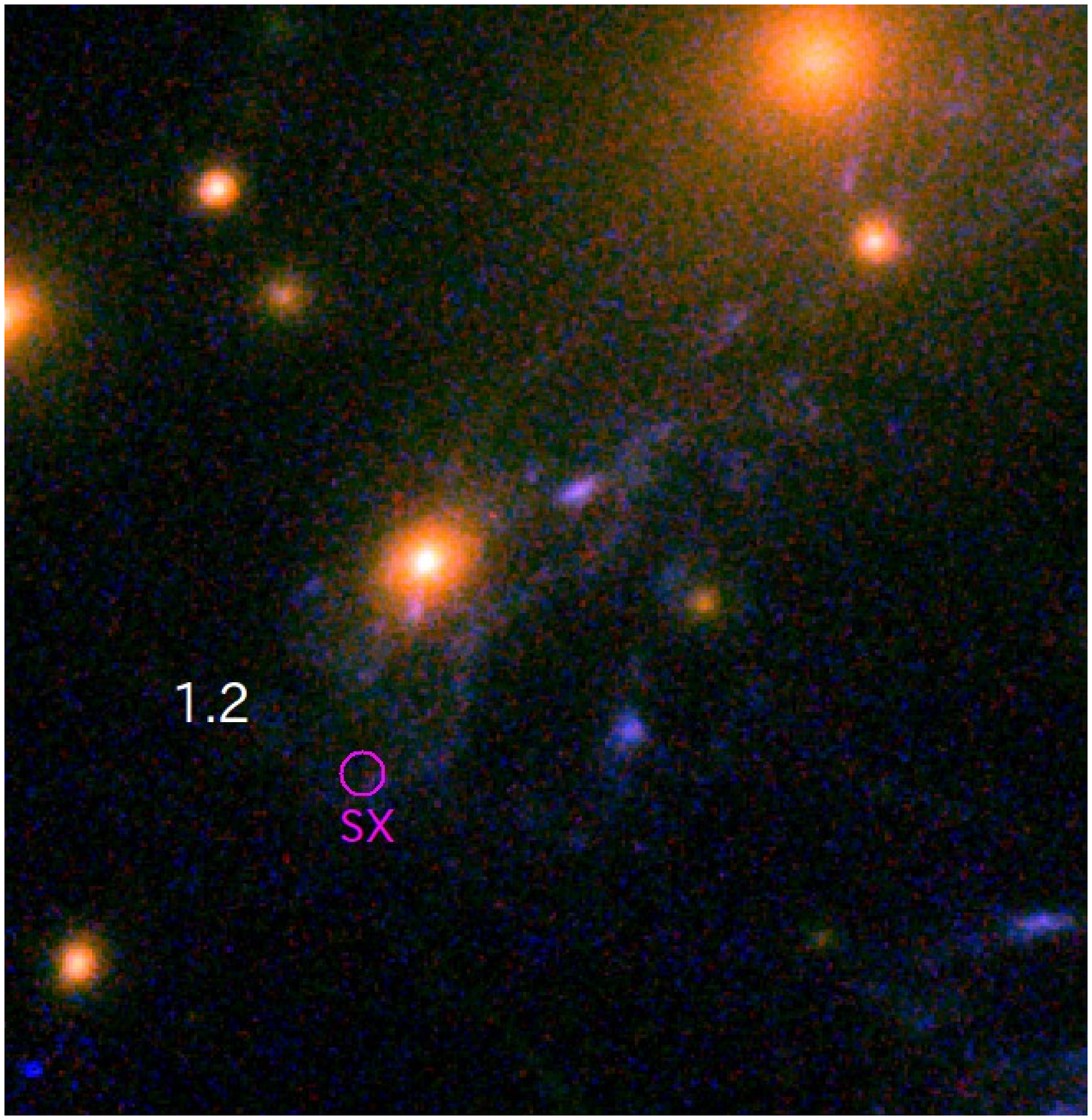}
 \includegraphics[width=0.32\hsize]{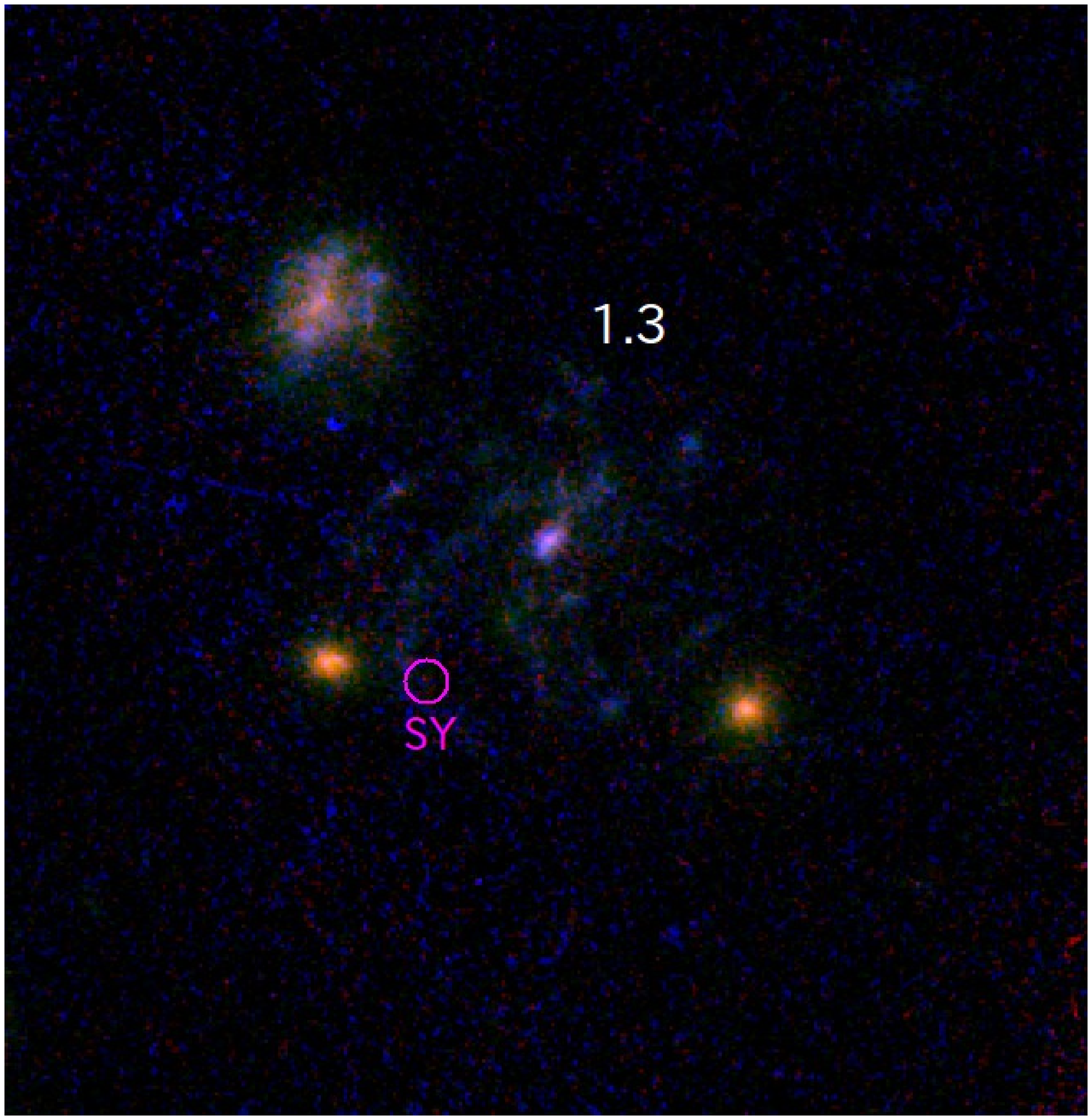}
\end{center}
\caption{Zoomed images of the regions 1.1, 1.2, and 1.3 (see
  Figure~\ref{fig:main}). Circles show the positions of six multiple
  images of SN Refsdal as predicted by our best-fit model. The
  properties are summarized in Table~\ref{tab:refsdal}.
\label{fig:zoom}}
\end{figure*}
%%%%%%%%%%%%%%%%%%%%%%%%%%%%%%%%%%%%%%%%%%%%%%%%%%%%%%%%%%%%%%%%%%%%%

%%%%%%%%%%%%%%%%%%%%%%%%%%%%%%%%%%%%%%%%%%%%%%%%%%%%%%%%%%%%%
%%%%%%%%%%%%%%%%%%%%%%%%%%%%%%%%%%%%%%%%%%%%%%%%%%%%%%%%%%%%%
\section{Result}\label{sec:result}
%%%%%%%%%%%%%%%%%%%%%%%%%%%%%%%%%%%%%%%%%%%%%%%%%%%%%%%%%%%%%
%%%%%%%%%%%%%%%%%%%%%%%%%%%%%%%%%%%%%%%%%%%%%%%%%%%%%%%%%%%%%

Our best-fit model produces $\chi^2=45$ for 40 degree of freedom,
 indicating that image positions predicted by the best-fit model
  agree with the observed positions with the root-mean-square value
  roughly equal to the input positional uncertainties. 
suggesting that the model reproduces multiple images almost with 
input positional uncertainties. Figure~\ref{fig:main} shows 
the critical curves at the redshift of the supernova, $z=1.491$,
predicted by our best-fit mass model. Our best-fit model predicts
six supernova images; in addition to the four Einstein cross images
S1--S4 (the region 1.1 in Figure~\ref{fig:main}) which was identified
by \citet{kelly15}, there is one extra image in the region 1.2, and
yet another extra image in the region 1.3. The predicted properties of 
the six supernova images are summarized in Table~\ref{tab:refsdal}. 

Our best-fit model predicts the high-magnifications of $\mu\sim 15-18$
for S1--S3, whereas the lower magnification of $\mu\sim 9$ for S4. The
time delay between S1--S4 are generally short. Time delays between
S1, S2, and S3 are all within 10 days, and the longest time delay
between any pairs of S1--S4 are the delay between S1 and S4, 22.5
days. These predictions are consistent with the current observational
data \citep{kelly15}.

An intriguing fact, which was also noted in \citet{kelly15}, is the
predicted presence of additional multiple images. We show the
predicted positions of all the six images in Figure~\ref{fig:zoom}.
SY is predicted to have appeared about 17 years ago, and hence there
is probably no chance of detecting this image in archival images,
unless the current time delay is over-predicted by a large amount. 
On the other hand, the image SX in the region 1.2 will appear in about
one year, and therefore this prediction can be tested in near future. 
We however note that the deep observations are required to detect SX
as the predicted magnification is smaller than that of S4. 

%%%%%%%%%%%%%%%%%%%%%%%%%%%%%%%%%%%%%%%%%%%%%%%%%%%%%%%%%%%%%%%%%%%%%
\begin{table}
 \caption{The properties of multiple images of SN Refsdal predicted by
   our best-fit model. $\mu$ denote the magnification
   factor, and  $\Delta t$ the time delay relative to S1.
\label{tab:refsdal}}    
 \begin{tabular}{@{}ccccc}
 \hline
   component
   & R.A.
   & Decl.
   & $\mu$
   & $\Delta t$ [days]
   \\
 \hline
S1 & 177.398224 &  22.395641 & 15.30 &     0.0 \\
S2 & 177.397721 &  22.395782 & 17.66 &     9.2 \\
S3 & 177.397383 &  22.395529 & 18.29 &     5.2 \\
S4 & 177.397799 &  22.395163 &  8.78 &    22.5 \\
SX & 177.399998 &  22.396723 &  3.94 &   357.1 \\
SY & 177.403749 &  22.402076 &  3.78 & $-6193.5$ \\
 \hline
 \end{tabular}
\end{table}
%%%%%%%%%%%%%%%%%%%%%%%%%%%%%%%%%%%%%%%%%%%%%%%%%%%%%%%%%%%%%%%%%%%%%%%

%%%%%%%%%%%%%%%%%%%%%%%%%%%%%%%%%%%%%%%%%%%%%%%%%%%%%%%%%%%%%
%%%%%%%%%%%%%%%%%%%%%%%%%%%%%%%%%%%%%%%%%%%%%%%%%%%%%%%%%%%%%
\section{Summary}\label{sec:summary}
%%%%%%%%%%%%%%%%%%%%%%%%%%%%%%%%%%%%%%%%%%%%%%%%%%%%%%%%%%%%%
%%%%%%%%%%%%%%%%%%%%%%%%%%%%%%%%%%%%%%%%%%%%%%%%%%%%%%%%%%%%%

In this paper we have constructed a mass model of MACS J1149.6+2223 to
predict the properties of multiple images of the recently discovered
strongly lensed supernova RN Refsdal at $z=1.491$ \citep{kelly15}. The
current data already provide reasonably good constraints on the mass
distribution of the core of the cluster. Our best-fit mass model
predicts that there are two additional images $\sim 10''$ and $30''$
away from the four Einstein cross images S1--S4 reported by
\citet{kelly15}. One image is predicted to have appeared about 17
years ago, whereas the other image will appear 1 year after the
appearance of S1--S4. The best-fit magnifications are $\sim 15$ for
S1--S3, $\sim 9$ for S4, and $\sim 4$ for the two extra images.  These
results should provide useful basis of the follow-up strategy of this
unique transient event. 

We note that the prediction is sensitive to the mass distribution of
the cluster \citep[see also][in which different time delays are
  predicted]{sharon15}, which suggests the importance of improving
mass models with ongoing and future observations. For instance, GLASS
observations, in which SN Refsdal was discovered, will provide
spectroscopic redshifts of many background galaxies behind MACS
J1149.6+2223, some of which should be multiply imaged
\citep[e.g.,][]{schmidt14}. Ongoing {\it Hubble} FF programmes also
provide much deeper {\it HST} images, from which many new sets of
multiple images will be discovered \citep[e.g.,][]{jauzac14}. These
new observations can be used to refine the mass model further, leading
to more robust predictions for the supernova images. On the other
hand, any mismatches between predicted and observed properties (e.g.,
time delays) indicate the room for improvements of our understanding
of cluster mass distributions. 

%%%%%%%%%%%%%%%%%%%%%%%%%%%%%%%%%%%%%%%%%%%%%%%%%%%%%%%%%%%%%
%%%%%%%%%%%%%%%%%%%%%%%%%%%%%%%%%%%%%%%%%%%%%%%%%%%%%%%%%%%%%
\section*{Acknowledgments}
%%%%%%%%%%%%%%%%%%%%%%%%%%%%%%%%%%%%%%%%%%%%%%%%%%%%%%%%%%%%%%
%%%%%%%%%%%%%%%%%%%%%%%%%%%%%%%%%%%%%%%%%%%%%%%%%%%%%%%%%%%%%%%
I thank R. Quimby, S. More, A. More, M. Werner, K. Nomoto for 
useful discussions. 
This work was supported in part by World Premier International
Research Center Initiative (WPI Initiative), MEXT, Japan, and
Grant-in-Aid for Scientific Research from the JSPS (26800093).

%%%%%%%%%%%%%%%%%%%%%%%%%%%%%%%%%%%%%%%%%%%%%%%%%%%%%%%%%%%%%%%%%%%%%%%

\label{lastpage}


\begin{thebibliography}{}

\bibitem[\protect\citeauthoryear{Amanullah et 
al.}{2011}]{amanullah11} Amanullah R., et al., 2011, ApJ, 742, L7 

\bibitem[\protect\citeauthoryear{Bertin \& Arnouts}{1996}]{bertin96}
Bertin E., Arnouts S., 1996, A\&AS, 117, 393 

\bibitem[\protect\citeauthoryear{Bolton 
\& Burles}{2003}]{bolton03}
Bolton A.~S., Burles S., 2003, ApJ, 592, 17 

\bibitem[\protect\citeauthoryear{Chornock et 
al.}{2013}]{chornock13} Chornock R., et al., 2013, ApJ, 767, 162 

\bibitem[\protect\citeauthoryear{Goobar et 
al.}{2002}]{goobar03}
Goobar A., M{\"o}rtsell E., Amanullah R., Nugent P., 2002, A\&A, 393, 25 

\bibitem[\protect\citeauthoryear{Holz}{2001}]{holz01} Holz 
D.~E., 2001, ApJ, 556, L71 

\bibitem[\protect\citeauthoryear{Ishigaki et 
al.}{2015}]{ishigaki15} Ishigaki M., Kawamata R., Ouchi M., Oguri 
M., Shimasaku K., Ono Y., 2015, ApJ, 799, 12 

\bibitem[\protect\citeauthoryear{Jauzac et al.}{2014}]{jauzac14} 
Jauzac M., et al., 2014, MNRAS, 443, 1549 

\bibitem[\protect\citeauthoryear{Johnson et 
al.}{2014}]{johnson14} Johnson T.~L., Sharon K., Bayliss M.~B., 
Gladders M.~D., Coe D., Ebeling H., 2014, ApJ, 797, 48 

\bibitem[\protect\citeauthoryear{Kelly et 
al.}{2015}]{kelly15}
Kelly P.~L., et al., 2015, arXiv:1411.6009

\bibitem[\protect\citeauthoryear{Kolatt 
\& Bartelmann}{1998}]{kolatt98}
Kolatt T.~S., Bartelmann M., 1998, MNRAS, 296, 763 

\bibitem[\protect\citeauthoryear{Navarro, Frenk, \& White}{1997}]{navarro97}
Navarro J.~F., Frenk C.~S., White S.~D.~M., 1997, ApJ, 490, 493 

\bibitem[\protect\citeauthoryear{Nordin et al.}{2014}]{nordin14} 
Nordin J., et al., 2014, MNRAS, 440, 2742 

\bibitem[\protect\citeauthoryear{Oguri}{2010}]{oguri10a}
Oguri M., 2010, PASJ, 62, 1017 

\bibitem[\protect\citeauthoryear{Oguri 
\& Kawano}{2003}]{oguri03b} Oguri M., Kawano Y., 2003, MNRAS, 338, L25 

\bibitem[\protect\citeauthoryear{Oguri \& Marshall}{2010}]{oguri10b}
Oguri M., Marshall P.~J., 2010, MNRAS, 405, 2579 

\bibitem[\protect\citeauthoryear{Oguri, Suto, 
\& Turner}{2003}]{oguri03a}
Oguri M., Suto Y., Turner E.~L., 2003, ApJ, 583, 584 

\bibitem[\protect\citeauthoryear{Patel et al.}{2014}]{patel14} 
Patel B., et al., 2014, ApJ, 786, 9 

\bibitem[\protect\citeauthoryear{Postman et 
al.}{2012}]{postman12} Postman M., et al., 2012, ApJS, 199, 25 

\bibitem[\protect\citeauthoryear{Quimby et al.}{2013}]{quimby13} 
Quimby R.~M., et al., 2013, ApJ, 768, L20 

\bibitem[\protect\citeauthoryear{Quimby et al.}{2014}]{quimby14} 
Quimby R.~M., et al., 2014, Sci, 344, 396 

\bibitem[\protect\citeauthoryear{Rau, Vegetti, 
\& White}{2014}]{rau14}
Rau S., Vegetti S., White S.~D.~M., 2014, MNRAS, 443, 957 

\bibitem[\protect\citeauthoryear{Refsdal}{1964}]{refsdal64} 
Refsdal S., 1964, MNRAS, 128, 307 

\bibitem[\protect\citeauthoryear{Richard et 
al.}{2014}]{richard14}
Richard J., et al., 2014, MNRAS, 444, 268 

\bibitem[\protect\citeauthoryear{Schmidt et 
al.}{2014}]{schmidt14} Schmidt K.~B., et al., 2014, ApJ, 782, 
L36 

\bibitem[\protect\citeauthoryear{Sharon 
\& Johnson}{2015}]{sharon15} 
Sharon K., Johnson T.~L., 2015, arXiv, arXiv:1411.6933 

\bibitem[\protect\citeauthoryear{Smith et al.}{2009}]{smith09} 
Smith G.~P., et al., 2009, ApJ, 707, L163 

\bibitem[\protect\citeauthoryear{Zitrin 
\& Broadhurst}{2009}]{zitrin09}
Zitrin A., Broadhurst T., 2009, ApJ, 703, L132 

\end{thebibliography}
\end{document}